\documentclass[aps,prb,reprint,twocolumn,showpacs,floatfix,superscriptaddress]{revtex4}
\usepackage{graphicx}
\usepackage{amssymb}
\usepackage{amsmath}
\usepackage{lmodern}
\usepackage{color}

\begin{document}

\title{Opening of a Gap in Graphene Due to Supercell Potential: Group Theory Point of View}

\author{E. Kogan}
\email{Eugene.Kogan@biu.ac.il}
\author{M. Kaveh}
\email{Moshe.Kaveh@biu.ac.il}
\affiliation{Jack and Pearl Resnick Institute, Department of Physics, Bar-Ilan University, Ramat-Gan 52900,
Israel}

\date{\today}

\begin{abstract}
We analyze in the framework of the space group theory the change of the dispersion law in grapenein
the vicinity of the (former) Dirac points due to application of supercell potential with the
$\sqrt{3}\times\sqrt{3}$ space periodicity and the same point symmetry as graphene.
\end{abstract}


\maketitle

Graphene is a two-dimensional crystal of carbon atoms, which form a honeycomb lattice with the point symmetry
described by the group $D_{6h}$. The first Brillouin zone (BZ) has a hexagonal form, and the conduction band
touches the valence band in six BZ corners which form two non-equivalent triads of BZ corners, K and K'.
One of the routes toward tailoring the electronic properties of graphene is through the adsorption of metals \cite{uchoa,giovannetti}.
Recently, several types of adatoms were used to dope graphene in attempts to tailor properties of graphenebased
devices \cite{elias,chen,zhou,schedin,lamoen}. The gap opening in the high symmetry points for the hexagonal lattices due to interaction
with the interface was considered in Ref. \onlinecite{vasseur}. In Ref. \onlinecite{farjam} was shown, using ab initio density functional  calculations,
that the adsorption of an alkali-metal submonolayer on graphene occupying every third hexagon of the
honeycomb lattice in a commensurate $(\sqrt{3}\times\sqrt{3})R30^{\circ}$ arrangement induces an energy gap in the spectrum of
graphene. We decided to analyze this opening of the gapin the framework of the group theory. In our previous
publications \cite{kogan1,kogan2,kogan3} we summed up the classification of the energy bands in graphene on the basis of the point
group analysis. The only fact from that analysis we need in the present work, is the fact that the little group at
the point K and is $D_{3h}$ and the bands $\pi$ and $\pi^*$ realize $E''$ representation of the group. The same
can be said about the point K'. This fact by itself means that the bands $\pi$ and $\pi^*$ touch each other at the
points K and K', and the electron states in the vicinity of these points are described by massless Dirac
equation \cite{kogan2}.

For the purpose of the present paper we should put the abovementioned fact into the framework of the theory of
the space group symmetry \cite{bradley,aroyo}, from which we will need only a few basic ideas. According to the theory of
the space group symmetry, the bands $\pi$ and $\pi^*$ should be considered at the points K and K' (these
two points can be considered as the stars of the wave vector ${\bf K}$, and designated $*{\bf K}$ together, thus realizing
a 4-dimensional representation of the $D_{6h}$ group. Due to the identity
\begin{eqnarray}
D_{6h}=D_{3h}\cup C_c\times D_{3h},
\end{eqnarray}
any element of the group 6h D can be presented as an element $G$ of the group $D_{3h}$, or as a product of $C_2$
and such element. Representation of the point group $D_{3h}$ $E''$ realized at the point K (K') defines representation
of the space group realized at $*{\bf K}$. The matrix representing an element G is a super-matrix $2\times 2$
\begin{eqnarray}
\label{space0}
D^{*{\bf K},E''}(G)=\left(\begin{array}{c|c} D^{E''}(G) & 0 \\\hline 0 & D^{E''}(G)\end{array}\right),
\end{eqnarray}
super-indices 1 and 2 referring to the points K and K' respectively. The matrix representing an element $C_2\times G$ is
\begin{eqnarray}
\label{space}
D^{*{\bf K},E''}(C_2\times G)=\left(\begin{array}{c|c} 0 & \dots\\\hline \dots  & 0\end{array}\right).
\end{eqnarray}
We will not need the exact form of the non-diagonal matrix elements in Eq. (\ref{space}); what we need is the fact
that the trace of the matrix $D^{*{\bf K},E''}(C_2\times G)$ is equal to zero. Naturally, when we consider dispersion in grapheme
as it is, space group symmetry point of view adds very little in comparison to point group symmetry
point of view, because the Hamiltonian, which has the symmetry $D_{3h}$, is block-diagonal.
Now we apply the group theory to analyze what happens at the points K, K' in grapheme with a perfectly
commensurate superlattice potential (which appears either because of the substrate or because of the absorbed
atoms), which has the same point symmetry 6h D as graphene. We consider explicitly a $\sqrt{3}\times\sqrt{3}$ superlattice,
known as the Kekule distortion of the honeycomb lattice \cite{farjam}. In this case we may consider the Brillouine zone
(BZ) of the superlattice as the folding of the original \cite{farjam,cheianov}. The folding leads to the identification of the corners
of the original BZ ( K and K' ) with the center $\tilde{\Gamma}$ of the new BZ. The Hamiltonian is no longer block
diagonal and, because the points K and K' are now identical, has the full symmetry $D_{6h}$. We thus observe
a paradox situation: due to decrease of the translational symmetry the point symmetry of the Hamiltonian has
increased.
Because of the symmetry of the Hamiltonian, we need to decompose representation realized by matrices (\ref{space0})
and (\ref{space}) with respect to the irreducible representations of the group 6h D .To obtain the decomposition, it is convenient
to use equation
\begin{eqnarray}
\label{ex}
a_{\alpha}=\frac{1}{g}\sum_G\chi(G)\chi_{\alpha}^*(G),
\end{eqnarray}
which shows how many times a given irreducible representation $\alpha$ is contained in a reducible one \cite{landau}.In Eq.
(\ref{ex}) $g$ is the number of elements in the group, $\chi_{\alpha}(G)$ is the character of an operator $G$ in the irreducible
representation $\alpha$ and $\chi(G)$ is the character of the operator G in the representation being decomposed.
Actually, even without using Equation (\ref{ex}), just by inspection of the two lowest line of Table 1 we obtain
the decomposition
\begin{eqnarray}
R=E_{1g}+E_{2u}.
\end{eqnarray}
We see that due to supercell potential two degenerate Dirac points disappear. At the point $\tilde{\Gamma}$ we have two
merging bands realizing representation $E_{1g}$ and another two merging bands realizing representation $E_{2u}$. We
may expect that representation $E_{1g}$ , as being more symmetrical, is realized at the top of the valence band,
and the representation $E_{2u}$ is realized at the bottom of the conduction band.
\begin{table}
\begin{tabular}{|l|l|rrrrrr|}
\hline
 $D_6$ &  & $E$ & $C_2$ & $2C_3$ & $2C_6$ & $3U_2$ & $3U_2'$ \\
 & $D_{3h}$   & $E$ & $\sigma$ & $2C_3$ & $2S_3$ & $3U_2$ & $3\sigma_v$ \\\hline
 $A_{1}$ & $A_1'$ & 1 & 1 & 1 & 1 & 1 & 1 \\
 $A_{2}$ & $A_2'$ & 1 & 1 & 1 & 1 & $-1$ & $-1$ \\
 $B_{1}$ & $A_1''$ & 1 & $-1$ & 1 & $-1$ & 1 & $-1$ \\
 $B_{2}$  &  $A_2''$ &1 & $-1$ & 1 & $-1$ & $-1$ & 1 \\
 $E_{2}$  & $E'$ & 2 & 2 & $-1$ & $-1$ & 0 & 0 \\
 $E_{1}$ & $E''$  & 2 & $-2$  & $-1$ & 1 & 0 & 0 \\
\hline
\end{tabular}
\caption{Character table for irreducible representations of  $D_6$ and $D_{3h}$ point groups}
\label{table:d2}
\end{table}

To get the form of the energy spectrum of the electrons in the vicinity of the point $\tilde{\Gamma}$ let us consider both the
${\bf k\cdot p}$ term \cite{pitaevskii} and the supercell potential as a perturbation. The effective Hamoltonian is
\begin{eqnarray}
\hat{H}_{eff}=\hat{V}+\hat{H}_{\bf k\cdot p},
\end{eqnarray}
where
\widetext
\begin{eqnarray}
\hat{H}_{\bf k\cdot p}=\left(\begin{array}{cccc} 0 & v(k_x-ik_y) & 0 & 0  \\ v(k_x+ik_y) & 0 & 0 & 0  \\ 0 & 0 & 0 &  v(k_x+ik_y)  \\ 0 & 0 &  v(k_x-ik_y) & 0  \end{array}\right),
\end{eqnarray}
\narrowtext
and $\hat{V}=\hat{V}^{(0)}$ reduces to two independent real constants
\begin{eqnarray}
\hat{V}^{(0)}=\left(\begin{array}{cccc} 0 & 0  & V_1 & 0 \\ 0 & 0 & 0 & V_2  \\ V_1 & 0 & 0 & 0   \\ 0 & V_2 & 0 & 0  \end{array}\right)
\end{eqnarray}
The specific form of the operator $\hat{V}^{(0)}$ follows from the symmetry of the base functions realizing representations
$E_{1g}$ and $E_{2u}$. By shifting origin of the energy axis these two constants can be chosen as
\begin{eqnarray}
V_1=-V_2=V.
\end{eqnarray}
Forming and solving the secular equation from these matrix elements, we obtain
\begin{eqnarray}
\epsilon^{(0)}(k)=\pm\sqrt{v^2k^2+V^2},
\end{eqnarray}
the sign plus corresponding to an upper pair of bands, and the sign minus corresponding to a lower pair of
bands.
To resolve between the branches in each pair we should take into account ${\bf k}$ corrections to the operator
$\hat{V}^{(0)}$. The first order in ${\bf k}$ corrections is equal to zero because the symmetry group contains the center of inversion.
To the second order in ${\bf k}$ we have ($i, j = 1,2$)
\begin{eqnarray}
\hat{V}=\hat{V}^{(0)}+\hat{V}^{(2)}=\hat{V}^{(0)}+\hat{\gamma}_{ij}k_ik_j,
\end{eqnarray}
where $\hat{\gamma}_{ij}$ is an Hermitian tensor operator (symmetrical in the suffixes $i$ and $j$ ). These include the corrections
from the terms linear in ${\bf k}$ in the Hamiltonian in the second-order perturbation theory and the corrections
from the terms quadratic in ${\bf k}$ in the first-order perturbation theory \cite{pitaevskii}. Notice that $\hat{V}^{(2)}$ is small relative to
both $\hat{V}^{(0)}$  and 
$\hat{H}_{\bf k\cdot p}$ (because we consider the states in the vicinity of the point $\tilde{\Gamma}$).
The relations exist between the matrix elements of the operator because of the requirements of symmetry.
As regards their transformation law under the symmetry operations, the wave functions which form the basis of
the representation $E_{2u}$  can be taken in the form
\begin{eqnarray}
\psi_1\sim zx,\;\;\;\psi_2\sim zy,
\end{eqnarray}
and the wave functions which form the basis of
the representation $E_{1g}$  can be taken in the form
\begin{eqnarray}
\psi_1\sim z,\;\;\;\psi_2\sim zxy.
\end{eqnarray}
From this, we easily conclude that in the first case the matrix elements of the $\hat{\gamma}_{ij}$ reduce to three independent
real constants
\begin{eqnarray}
<1|\gamma_{xx}|1>=<2|\gamma_{yy}|2>=A\nonumber\\
<2|\gamma_{xx}|2>=<1|\gamma_{yy}|1>=B\nonumber\\
<1|\gamma_{xy}|2>=<2|\gamma_{xy}|1>=C.
\end{eqnarray}
The matrix elements of the operator $\hat{V}^{(2)}$ are
\begin{eqnarray}
<1|\hat{V}^{(2)}|1>=<2|\hat{V}^{(2)}|2>=Ak_x^2+Bk_y^2\nonumber\\
<1|\hat{V}^{(2)}|2>=<2|\hat{V}^{(2)}|1>=2Ck_xk_y.
\end{eqnarray}
In the second case the matrix elements of the $\hat{\gamma}_{ij}$  also reduce to three independent real constants
\begin{eqnarray}
<1|\gamma_{xx}|1>=<1|\gamma_{yy}|1>=D\nonumber\\
<2|\gamma_{xx}|2>=<2|\gamma_{yy}|2>=E\nonumber\\
<1|\gamma_{xx}|2>=<1|\gamma_{yy}|2>=F.
\end{eqnarray}
The matrix elements of the operator $\hat{V}^{(2)}$ are
\begin{eqnarray}
<1|\hat{V}^{(2)}|1>&=&Dk^2\nonumber\\
<2|\hat{V}^{(2)}|2>&=&Ek^2\nonumber\\
<1|\hat{V}^{(2)}|2>&=&2Fk_xk_y.
\end{eqnarray}

Forming and solving the secular equation from these matrix elements, we obtain for the $E_{1g}$ branches of
the spectrum
\begin{eqnarray}
\epsilon({\bf k})=\epsilon^{(0)}(k)+Ak_x^2+Bk_y^2\pm2Ck_xk_y.
\end{eqnarray}
The formula for the $E_{2u}$ branches of the spectrum can be obtained similarly.

The folding of the BZ, together with the destruction of previously existing gapless Dirac points, leads to appearance
of the new ones. In fact, the new BZ is still a hexagon, and the same symmetry arguments used for
graphene can be used to explain appearance of the gapless Dirac points at the corners of the new BZ ($\tilde{K},\tilde{K}'$).
However, these new Dirac points are situated deep below or high above the Fermi level and, hence,  manifest
themselves less than Dirac points of unreconstructed graphene

\end{document}